\documentclass{PoS}
\usepackage{subcaption}
\usepackage{lineno}

\title{Measurement of longitudinal single-spin asymmetries for $W^{\pm}$ boson production in polarized $p+p$ collisions at $\sqrt{s}=510$ GeV at STAR}			

\ShortTitle{$W^{\pm}$ $A_L$ Measurement at STAR}

\author{\speaker{Devika Gunarathne} {(for the STAR collaboration)}\\
        Temple University, Philadelphia, PA, USA\\
        E-mail: \email{devika@temple.edu}}

\abstract{
$W^\pm$ boson production in longitudinally polarized $p+p$ collisions provides unique and clean access to the individual helicity polarizations of $u$ / $d$ quarks and anti-quarks. Due to the maximal violation of parity in the coupling, $W$ bosons couple to left-handed quarks and right-handed anti-quarks and hence offer direct probes of their respective helicity distributions in the nucleon.  These can be extracted from measured parity-violating longitudinal single-spin asymmetries, $A_L$, for $W^{+(-)}$ boson production as a function of the decay lepton (positron) pseudo-rapidity $\eta$. The STAR experiment is well equipped to measure $A_L$ for $W^\pm$ boson production for $|\eta|<1$. 
The published STAR $A_L$ results (2011 and 2012 data combined) have been used by several theoretical analyses suggesting a significant impact in constraining the helicity distributions  of anti-$u$ and anti-$d$ quarks. 
In 2013 the STAR experiment has collected a large data sample of $\sim$250 pb$^{-1}$ which is  more than 3 times larger than the total integrated luminosity in 2012, at $\sqrt{s}=510$ GeV with an average beam polarization of $\sim$54\%, comparable to run 2012. The status of the 2013 $A_L$ analysis will be discussed along with an overview of future plans. 
 }

\FullConference{The XXIII International Workshop on Deep Inelastic Scattering and Related Subjects\\
		April 27 - May 1, 2015\\
                Southern Methodist University\\
		Dallas, Texas 75275}

\begin{document}

\section{Introduction}
There has been steady progress over the past few decades in terms of understanding  the spin structure of the nucleon, one of the fundamental questions in nuclear physics. In the 1980s, the spin of the proton was naively explained~\cite{dis:rqpModel} by the alignment of spins of the valence quarks. However, in our current understanding~\cite{dis:EMU} , the valence quarks, sea quarks, gluons and their possible orbital angular momentum are all expected to contribute to the overall spin of the proton. Despite this significant progress, our understanding of the individual polarizations of quarks and antiquarks is not yet complete.
 	According to the spin sum rule introduced by Jeffe and Monahar~\cite{dis:sumrule} in 1990, the spin of the proton can be written in terms of its contributions from the intrinsic quark and antiquark polarization, intrinsic gluon polarization and their possible orbital angular momentum.  Polarized inclusive deep-inelastic scattering (DIS) experiments were able to strongly constrain the total quark contribution to the proton spin~\cite{dis:global}. However, DIS experiments were not sensitive to the flavor separated individual quark spin contributions. These were then measured by polarized semi inclusive DIS experiments (SIDIS), but relatively large uncertainties were observed in the extracted helicity-dependent parton distribution functions (PDF)~\cite{dis:global} of antiquarks compare to quarks. However, this method is limited by uncertainties in the fragmentation process~\cite{dis:frag}. The production of $W^\pm$ bosons in longitudinally polarized $p+p$ collisions at RHIC provides a unique and powerful tool to probe the individual helicity PDFs of light quarks and anti-quarks in the proton. Due to the maximal  parity violating nature of the weak interaction, $W^{-(+)}$ bosons couple to the left-handed quarks and right-handed anti-quarks and hence offer direct probes of their respective helicity distributions in the nucleon. These distributions can be extracted by measuring the parity-violating longitudinal single-spin asymmetry, $A_L$, as a function of the decay electron (positron) pseudo-rapidity, $\eta_e$. The longitudinal single-spin asymmetry is defined as $A_L =(\sigma_+-\sigma_- )/ (\sigma_++\sigma_-)$, where $\sigma_{+(-)}$ is the cross section when the  helicity of the polarized proton beam is positive (negative).
At leading order, $W^+$ $A_L$ is directly related to polarized anti d and u  quark distributions ($\Delta \bar d$, $\Delta u$) while  $W^-$$A_L$ is directly related to polarized anti u and d  quark distributions ($\Delta \bar u$, $\Delta d$)~\cite{dis:rhicW}. 

\par
The results~\cite{dis:run12paper} of the single-spin asymmetries for $W^\pm$ boson production in longitudinally polarized $p+p$ collisions from the 2011 and 2012 STAR running periods are presented. The integrated luminosity of the data set collected during these two years were 9 and 77 $pb^{-1}$, with an average beam polarization of $49\%$ and $56\%$, respectively. In 2013 the STAR experiment has collected a data sample of $\sim$250 pb$^{-1}$ at $\sqrt{s}=510$ GeV with an average beam polarization of $\sim$54\%.  The status of the 2013 W $A_L$ analysis is discussed. 

\section{Analysis}
The STAR experiment~\cite{dis:starNIM} is well equipped to measure $A_L$ for $W^\pm$ boson production within a pseudorapidity range of $|\eta|<1$. $W^\pm$ bosons are detected via their $W^\pm \rightarrow e^\pm \nu$ decay channels. A subsystem of the STAR detector, the Time Projection Chamber (TPC) is used to measure the transverse momentum ($p_T$) of decay electrons and positrons and to separate their charge sign. Two other subsystems, Barrel and Endcap Electromagnetic Calorimeters (BEMC, EEMC) are  used to measure the energy of decay leptons. A well developed algorithm~\cite{dis:run12paper} is used to identify and reconstruct $W^\pm$ candidate events by reducing large QCD type background events. In this algorithm, various cuts are designed at each level of the selection process based on the kinematics and topological differences between the electroweak process of interest and QCD processes. For example, tracks associated with $W^\pm$ candidate events can be identified as isolated tracks in the TPC that point to an isolated EMC cluster in the calorimeter, where as for QCD type events have several TPC tracks point to several EMC clusters. In contrast to QCD background events, large opposite missing transverse energy can be observed in $W^\pm\rightarrow e^\pm \nu$ decay, due to undetected neutrinos. This leads to a large imbalance in the vector $p_T$ sum of all reconstructed final-state objects in W candidate events, which is expressed as $\vec{p}^{balance}_T$ in equation 2.1. Here $\vec{p}^{jets}_T$ is determined using the anti-$\it{k}_T$ algorithm~\cite{dis:rhicW}. A cone of radius of 0.7 in $\eta-\phi$ space is centered around the candidate lepton, and the $p_T$ from all reconstructed jets outside of the cone are included. The $\vec{p}^{e}_T$ is the $p_T$ of the candidate lepton. A strong correlation can be observed between $E_T$ and scalar quantity, signed $p_T$ balance, which is defined in equation 2.2. This can be seen clearly in Figure 1 c) which shows the MC results simulating $W^\pm\rightarrow e\nu$ decays. After initially requiring reconstructed TPC tracks to have $p_T$ > 10 GeV, tracks are matched to a $2x2$ EMC cluster with transverse energy $E_T$ > 12 GeV. W candidate events are then isolated using the large opposite missing energy requirement. This is done by requiring that the $E_T$ fraction of the $2x2/4x4$ tower clusters is larger than $95\%$. The final requirement, which is based on the large imbalance in the vector $p_T$ sum mentioned above, is signed-$p_T$ balance to be greater than 14 GeV, which is indicated by the red line in Figure 1 b). After all the selection cuts have been applied, the characteristic Jacobian peak in the $E_T$ distribution for mid-rapidity $W^\pm$ candidate events can be observed near half of the $W^\pm$ mass, as shown in Figure 1 (a).

\begin{equation}
\vec{p}^{balance}_T=\vec{p}^{e}_T+\sum\limits_{\Delta R > 0.7} \vec{p}^{jets}_T
\end{equation}
\begin{equation}
signed\:{p_T}\:balance=\frac{(\vec{p}^e_T).\vec{p}^{balance}_T}{|\vec{p}^e_T|}
\end{equation}
 
\begin{figure}[!h]
\centering
\begin{subfigure}[b]{0.49\textwidth}
\includegraphics[width=\textwidth]{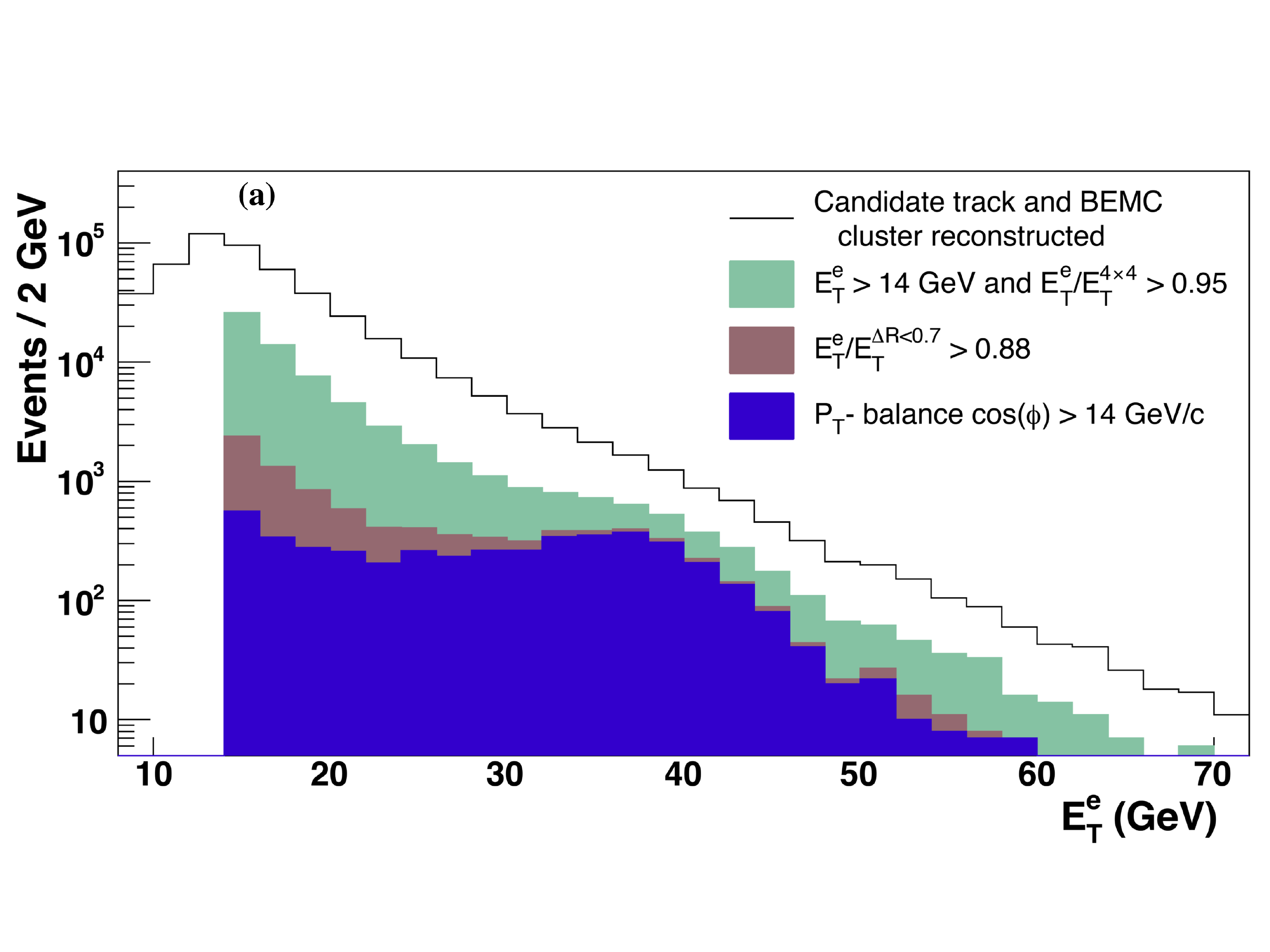}
\label{cuts}
\end{subfigure}
\begin{subfigure}[b]{0.49\textwidth}
\includegraphics[width=\textwidth]{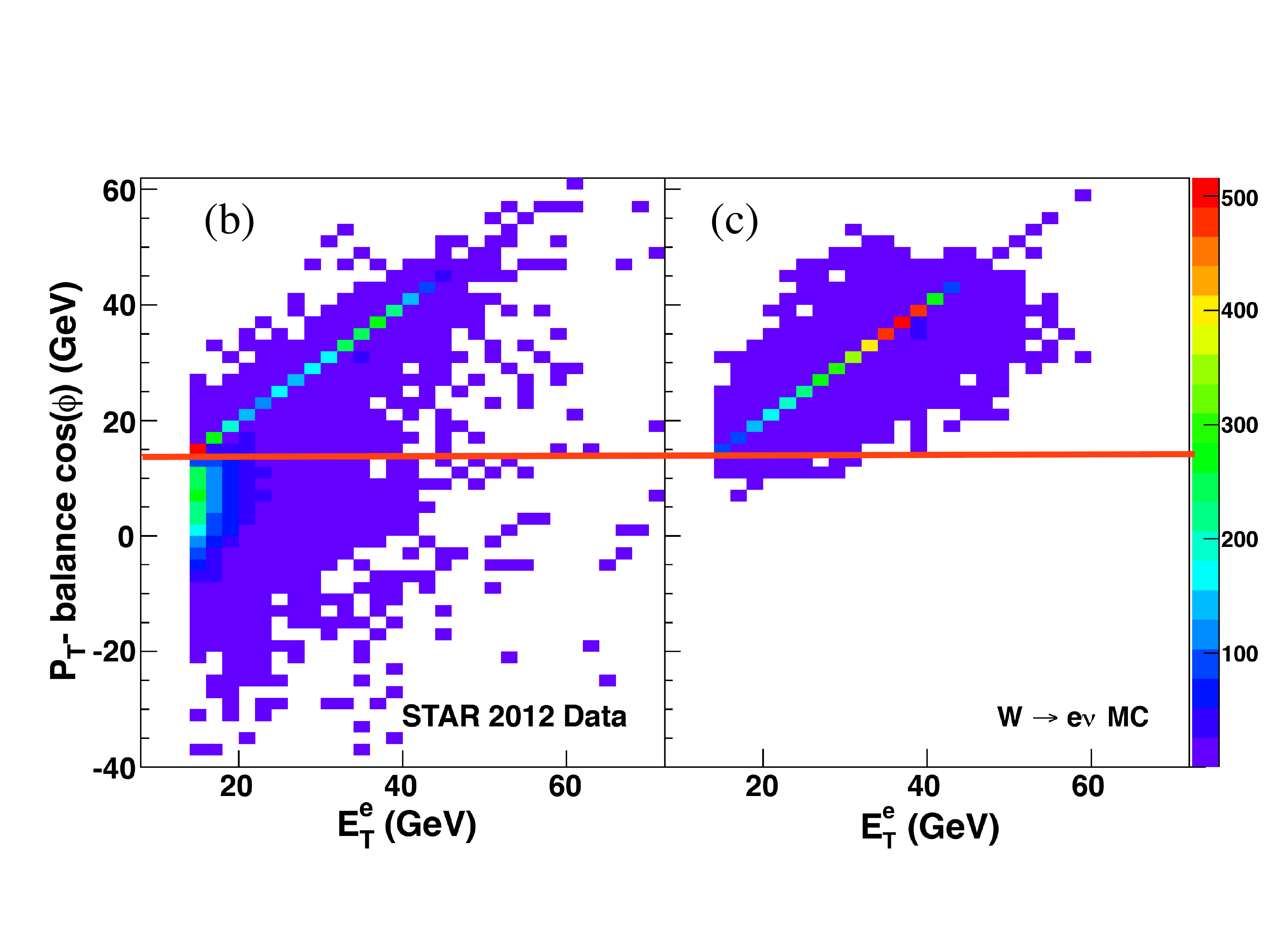}
\label{fig:signpt }
\end{subfigure}
\caption{Candidate $E^e_T$ distribution from the data after various selection cuts (a), Signed $p_T$-balance vs $E^e_T$ for data (b) and $W \rightarrow e\nu$ MC (c).~\cite{dis:run12paper}}\label{fig:cutsandsignpt}
\label{fig:2images}
\end{figure}

The Charge separated $W^\pm$ yields from the 2011 and 2012 data sets as a function of $E^e_T$ are shown in Figure 2 for different $\eta$ bins, along with the estimated residual background contributions from $W^\pm\rightarrow\tau^\pm\nu_\tau$,  $Z/{\gamma^*}\rightarrow e^+ e^-$ electroweak processes and QCD processes. Relatively small electroweak background contributions are estimated from Monte-Carlo (MC) simulation, with PYTHIA 6.422~\cite{dis:pythia} generated events passing through the STAR GEANT\cite{dis:geant} model and embedded in to STAR zero-bias triggered events.
Despite a significant reduction of QCD background events during the selection process, a certain amount is still present in the signal region. This contribution originates primarily from events which satisfy candidate $W^\pm$ isolation cuts but contain jets which escape the detection outside the STAR acceptance. Two procedures referred as "Second EEMC" and "Data-driven QCD"~\cite{dis:algo}, are used to estimated the background associated with the acceptance ranges $-2<\eta < -1.09$ and $|\eta|>2$. 
At forward rapidity, ($1<\eta_e <1.4$), the W selection criterion used is similar to that of mid rapidity. The background estimation is improved using additional Endcap Shower Maximum Detector (ESMD). More details are described in~\cite{dis:run12paper}.

\begin{figure}[!h]
\centering
\includegraphics[width=0.8\textwidth]{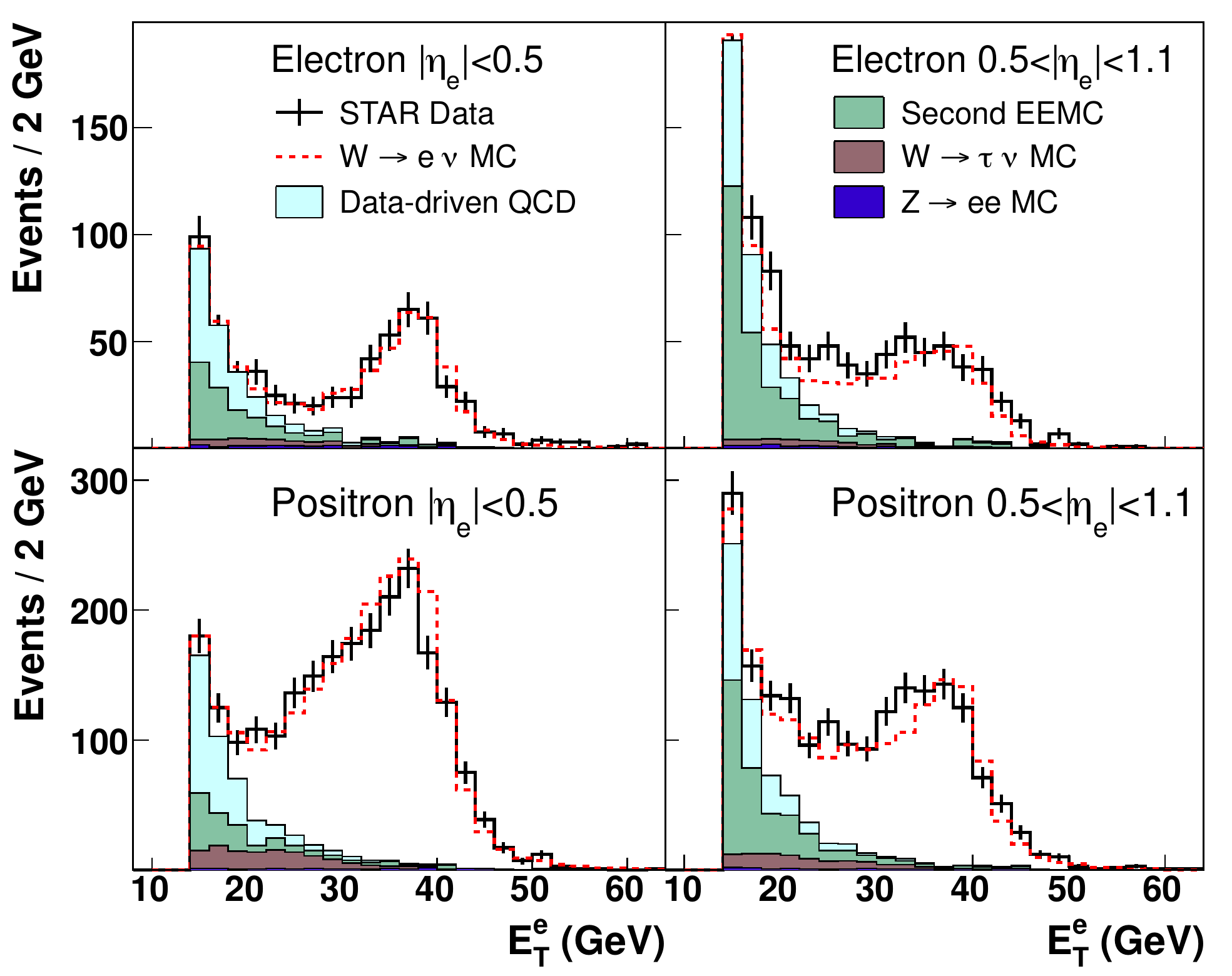}
\label{fig:bg}
\caption{$E^e_T$ distribution of $W^-$ (top) and $W^+$ (bottom) candidate events (black), various background contributions and sum of backgrounds and $W\rightarrow e\nu$ MC signal (red-dashed).~\cite{dis:run12paper}}
\end{figure}

\section{Results}
To properly account for the low statistics in the 2011 data set, a profile likelihood method was used to extract the spin asymmetry results from the combined 2011 and 2012 data sets. Two likelihood functions ${\it L}_{year1}$ and ${\it L}_{year2}$ are defined for each of the 2011 and 2012 data sets respectively. The asymmetry results, central values and confidence intervals are extracted from the product of the likelihood function,  ${\it L}_{2011} \times {\it  L}_{2012}$. More details are described in \cite{dis:run12paper}. The $W^\pm$ single-spin asymmetry results measured for $e^\pm$ with $25<E^e_T<50 GeV$ are shown in Figure 3 as the function of decay $e^\pm$ pseudorapidity, $\eta_e$ in comparison to theoretical predictions based on DSSV08~\cite{dis:dssv08} and LSS10 ~\cite{dis:lss10} helicity-dependent PDF sets, using both CHE (next-to-leading order)~\cite{dis:rhicW}  and RHICBOS (fully resummed) frameworks~\cite{dis:rhicbos}. The measured $A^{W^-}_L$ is larger than the central value of the theoretical predictions. The enhancement at large negative $\eta_e$, in particular is sensitive to the polarized anti u quark distribution, $\Delta\bar u$. $A^{W^+}_L$ is negative as expected and consistent with theoretical predictions. The systematic uncertainties for $A^{W^\pm}_L$ are well under control for pseudorapidity range $|\eta_e|<1.4$. STAR 2012 preliminary $A^{W^\pm}_L$ results~\cite{dis:global} are included in the DSSV++ global analysis~\cite{dis:dssv++} from the DSSV group and recent NNPDF~\cite{dis:nnpdf} global analysis. Both analyses show that the STAR W $A_L$ results provide a significant constraint on anti u ($\Delta\bar u$) and anti d ($\Delta\bar d$) quark polarizations.
 
\begin{figure}[!h]
\centering
\includegraphics[width=0.65\textwidth]{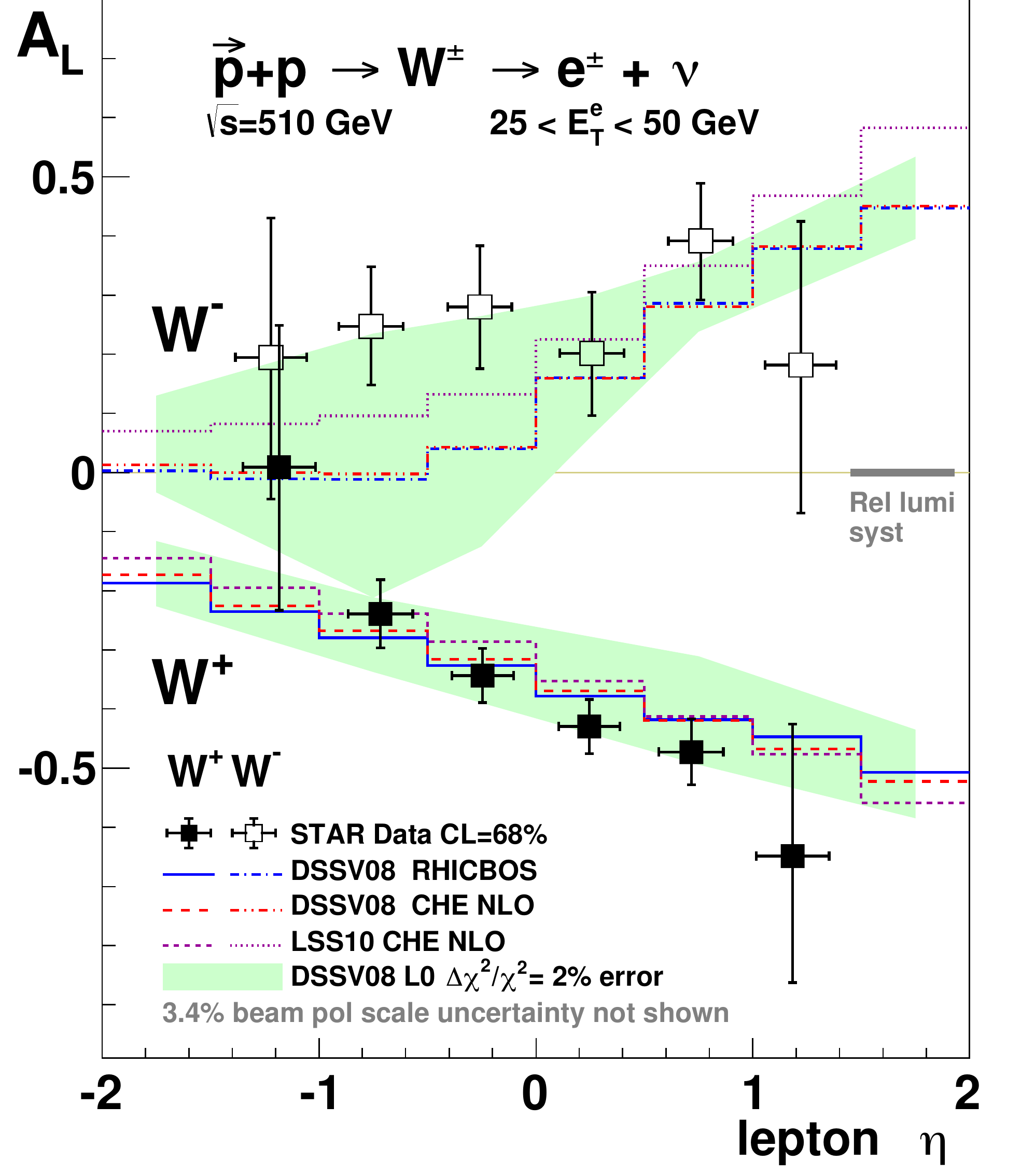}
\label{fig:results}
\caption{Longitudinal single-spin asymmetries for $W^\pm$ production as a function of lepton pseudorapidity, $\eta_e$ in comparison to theory predictions.~\cite{dis:run12paper} }
\end{figure}

\section{Outlook}
In 2013, the STAR experiment collected a large data sample of $\sim$250 pb$^{-1}$, which is  more than 3 times larger than the total integrated luminosity in 2012, at $\sqrt{s}=510$ GeV with an average beam polarization of $\sim$54\%. The high luminosity data collected in 2013 require proper calibration of all the subsystems used in the analysis. As the charge sign reconstruction of $e^+$ and $e^-$ is based on the bending of TPC tracks in the presence of an axial magnetic field, the calibration of the TPC is crucial for the W analysis. Despite the challenging environment in the calibration process due to large pile up accumulated in the TPC due to high luminosity $p+p$ collisions in 2013, a clear separation between  $e^+$ and $e^-$ is observed. 
The calibration of the other crucial subsystem for the mid-rapidity W analysis, the BEMC, is currently in progress. Forward Gem Tracker (FGT) was fully installed at STAR  in year 2013, which covers the acceptance between $1<\eta<2$ in the forward region. The FGT will be used as the tracking device for W analysis in the forward pseudorapidity region at STAR. This enhances the sensitivity to $\bar u$ and $\bar d$ quark polarizations. Figure 4 shows the projected uncertainties for the $W^\pm$ asymmetries estimated from the 2013 data set.

\begin{figure}[!h]
\centering
\includegraphics[width=1.0\textwidth]{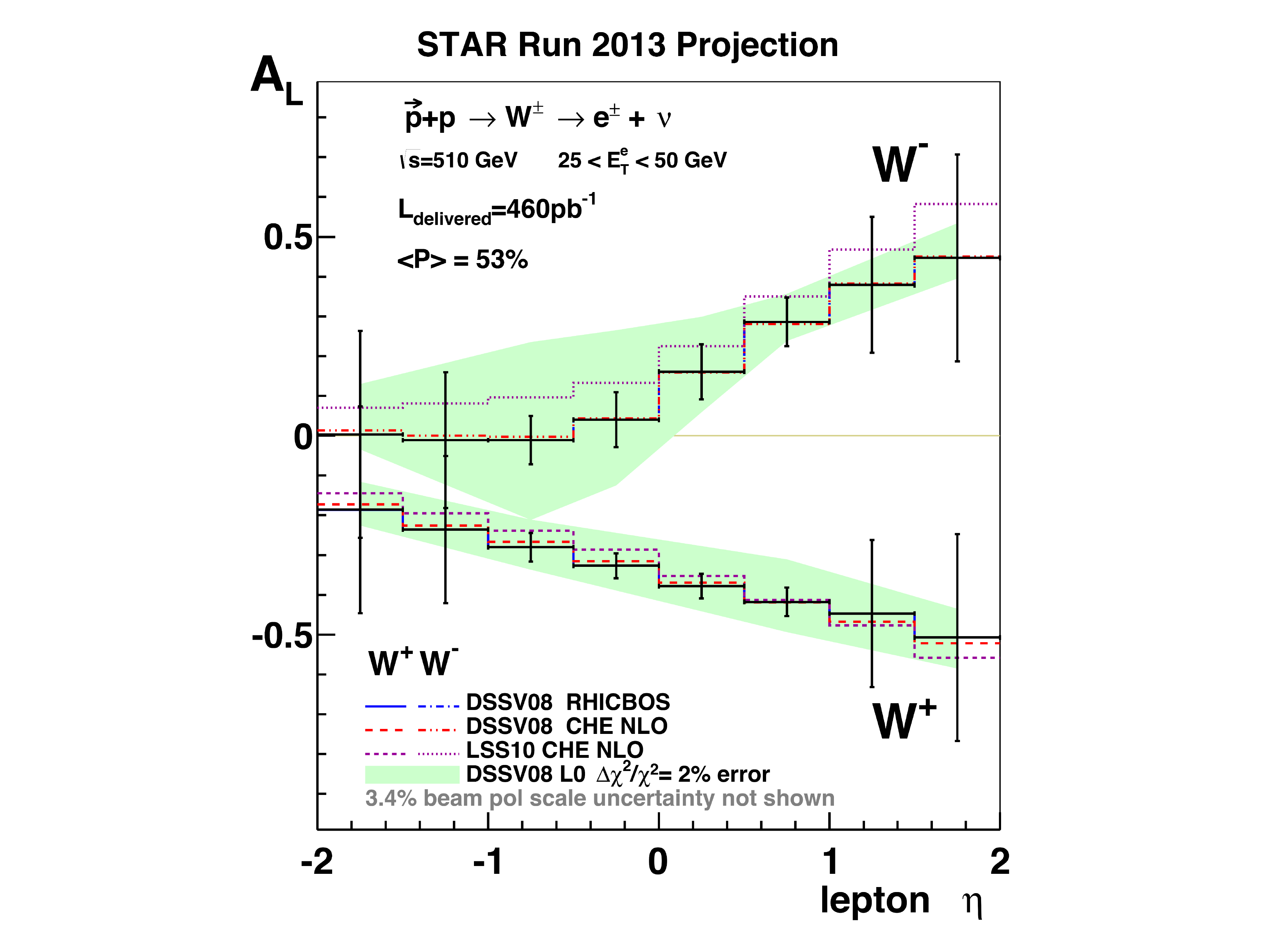}
\label{fig:projec}
\caption{Projected uncertainties for the longitudinal singal-spin asymmetries as a function of lepton pseudorapidity, $\eta_e$ for $W^\pm$ production from the 2013 data set. }
\end{figure}
Higher precision results are expected from the STAR 2013 W $A_L$ analysis to improve constraints on the sea quarks helicity-dependent PDFs.

\end{document}